# Relativistic Mott transition in strongly correlated artificial graphene

Liguo Ma[1,2], Raghav Chaturvedi[1], Phuong X. Nguyen[1,3], Kenji Watanabe[4], Takashi Taniguchi[4], Kin Fai Mak[1,3,5*], Jie Shan[1,3,5*]

[1]School of Applied and Engineering Physics, Cornell University, Ithaca, NY, USA
[2]Zhejiang Key Laboratory of Micro-nano Quantum Chips and Quantum Control and School of Physics, Zhejiang University, Hangzhou, China
[3]Kavli Institute at Cornell for Nanoscale Science, Ithaca, NY, USA
[4]National Institute for Materials Science, Tsukuba, Japan
[5]Laboratory of Atomic and Solid State Physics, Cornell University, Ithaca, NY, USA

*Email: jie.shan@cornell.edu; kinfai.mak@cornell.edu
These authors contributed equally: Liguo Ma, Raghav Chaturvedi, Phuong X. Nguyen

**The realization of graphene has provided a bench-top laboratory for quantum electrodynamics[1,2]. The low-energy excitations of graphene are two-dimensional massless Dirac fermions with opposite chiralities at the $\pm$K valleys of the graphene Brillouin zone[3]. It has been speculated that the electron-electron interactions in graphene could spontaneously break the chiral symmetry to induce a finite mass for Dirac fermions, in analogue to dynamical mass generation in elementary particles[4-15]. The phenomenon is also known as the relativistic Mott transition[6,8,12,14] and has not been observed in pristine graphene because the interaction strength is insufficient. Here, we report the realization of strongly correlated artificial graphene and the observation of the relativistic Mott transition in twisted $WSe_2$ tetralayers. Using magneto transport, we show that the first Γ-valley moiré valence band mimics the low-energy graphene band structure. At half-band filling, the system exhibits hallmarks of massless Dirac fermions, including an anomalous Landau fan originated from a π-Berry phase and a square-root density dependence of the cyclotron mass[1,2]. We tune the interaction across the semimetal-insulator transition by reducing the twist angle below about 2.7 degrees. The emergent insulator is compatible with an antiferromagnetic Mott insulator[6,8,12,14]. Our results open the possibility of studying strongly correlated Dirac fermions in a condensed matter system.**

## Main

Graphene's unique electronic properties originate from its two-dimensional (2D) honeycomb lattice structure[3]. The lattice symmetry dictates two gapless points at the opposite corners ($\pm$K) of the graphene Brillouin zone, around which the electronic bands are cone-shaped and are composed of states belonging to two different sublattices of the honeycomb. Low-energy electrons in graphene behave like chiral massless Dirac fermions in quantum electrodynamics (QED) with their pseudospin--index of the two sublattices--playing the role of spin and locked to the momentum. Electrons at the $\pm$K valleys have opposite chiralities. Breaking the chiral symmetry, namely, direct coupling states from different sublattices, is expected to open an energy gap at the Dirac point proportional to the newly generated electron mass[6-15]. Many interesting QED phenomena, such as Klein tunneling[16], a zeroth Landau level[1-3] and the Schwinger effect[17], have been demonstrated in graphene. But spontaneous chiral symmetry breaking driven by electron-electron interactions[6-15], also known as the relativistic Mott transition, has not been observed in pristine graphene presumably due to insufficient interaction strength. Only a Kekulé bond order has been reported in graphene doped by surface adatoms[18,19].

The interaction strength of massless Dirac fermions can be quantified by the fine-structure constant[20]. In graphene, the electron Fermi velocity $v_F$ plays the role of the speed of light $c$, thereby enhancing the fine-structure constant by a factor of $c/v_F \approx 300$. Further slowing down the Fermi velocity is required to access the relativistic Mott transition. The Fermi velocity is given by $v_F = \frac{\sqrt{3}ta}{2\hbar}$ in the tight binding model[3], where $t$ is the nearest-neighbor hopping integral, $a$ is the lattice constant and $\hbar$ is the Planck's constant. Increasing the lattice constant has been proposed as an efficient route to reducing the Fermi velocity[21,22] since the hopping integral decreases exponentially with lattice constant. The newly emerged moiré superlattices formed by stacking 2D materials with a small relative twist angle[23-27] provide an ideal platform to realize artificial graphene with highly tunable fine-structure constant[22,28]. In contrast to atomic lattices, the period of moiré superlattices can be continuously tuned over a wide range through the twist angle.

Here, we demonstrate that the first Γ-valley moiré valence band in twisted $WSe_2$ tetralayers simulates physics of strongly correlated graphene (single orbital on a honeycomb lattice). The Γ-valley states with negligible spin-orbit coupling can be described by a single-band Hamiltonian[22,28]. But in $WSe_2$ monolayers and bilayers, the valence band maximum is located at the K points with large spin-orbit coupling and minimal interlayer coupling[29]. We choose $WSe_2$ tetralayers in which the valence band maximum is located at the Γ point[28,30]. The Γ-valley states are mainly made of out-of-plane atomic orbitals and are pushed above the K valley by the strong interlayer hybridization[31]. The moiré superlattice is formed at the interface between two natural $WSe_2$ bilayers (of AB structure) with a small angle from 180° twist and is referred to as 'AB-BA' $WSe_2$ below (in contrast to the 'AB-AB' structure examined in Ref. [30]). It has been shown that the moiré potential for holes has $D_6$ symmetry with minima located at the MX and XM sites (M = W; X = Se), forming a honeycomb lattice[22,28] (Fig. 1a). The twisted tetralayers are also mechanically more robust and have less moiré disorders than twisted bilayers. The current experiment contrasts the existing ones on $WSe_2$ moiré materials that have largely focused on the K-valley holes[27,32-38] simulating either a single-orbital model on a triangular lattice[39] or a two-orbital model on a honeycomb lattice[40,41].

Figure 1b illustrates the electronic band structure of 3°-twisted AB-BA $WSe_2$ calculated following the reported continuum model[28]. The first moiré valence band is graphene-like with linear band crossing at the $\pm\kappa$ points and saddle points at the $m$ point of the moiré Brillouin zone. Hole doping density of $\nu = 2$, in units of moiré density $n_M$, is required to bring the Fermi level (red dotted line) to the Dirac point. Figure 1c shows the corresponding density of states (DOS) as a function of $\nu$. At the Dirac point the DOS vanishes. The saddle points give rise to two van Hove singularities at $\nu \approx 1.5$ and 2.3, which are close to the values, 3/2 and 5/2, predicted by the tight-binding model with nearest-neighbor hopping[3]. The graphene-like band structure is independent of twist angle and lattice relaxation due to the $D_6$ symmetry of the moiré lattice (Extended Data Fig. 2). With decreasing twist angle, the bandwidth or hopping integral decreases rapidly, supporting stronger interaction effects.

**Γ-valley moiré bands**

We perform electrical transport measurements on dual-gated AB-BA WSe$_2$ samples with twist angle $\theta$ ranging from 1.9° to 3.6° (Fig. 1a). The twist angle is calibrated from the Landau levels (Methods). The two gates (TG and BG) allow independent control of the hole doping density or the lattice filling factor, $\nu$, and the electric field, $E$, perpendicular to the sample plane. To achieve low contact resistances, we use Pt as contact electrodes and pattern contact gates (CG) to heavily hole dope the contact regions of AB-BA WSe$_2$. Unless otherwise specified, results are shown for lattice temperature $T = 20$ mK. See Methods for details on the device fabrication and electrical transport measurements.

Figure 1d displays the longitudinal resistance $R_{xx}$ of a 3°-twisted AB-BA WSe$_2$ sample as a function of $\nu$ and $E$. ($R_{xx}$ is approximately the resistivity given the device dimension shown in Extended Data Fig. 1.) Results for other twist angle samples are included in Extended Data Fig. 3. The phase diagram in Fig. 1d can be divided into three doping regions corresponding to the Γ, Γ+K and K valleys in WSe$_2$, respectively. Under small electric fields between the two black dashed lines, holes are doped into the Γ valley only. We observe several resistance peaks, the nature of which will be discussed later. Under high electric fields above the orange dashed lines, the K valley is pushed above the Γ valley by the Stark effect, and holes are doped into the K valley only. Between the two regions, the K and Γ valleys are comparable in energy, and both valleys are doped. The boundaries of these regions are identified from the electric-field dependence of the Landau levels (Extended Data Fig. 4). The Landau levels do not disperse with electric field when only the K or the Γ valley is doped. But in the Γ+K region, two sets of Landau levels are discernable, and they disperse with electric field because the electric field varies the partition of the doping density in different valleys.

The above assignment is further supported by the in-plane magnetic-field ( $B_\parallel$ ) dependence of $R_{xx}$ in Fig. 1e. For a representative doping density ($\nu \approx 1.2$), $R_{xx}$ shows a negligible dependence on $B_\parallel$ for the K-valley holes because their spin is pinned to the out-of-plane direction by the large Ising spin-orbit coupling[42]. In contrast, $R_{xx}$ strongly depends on $B_\parallel$ for the Γ-valley holes because they have negligible spin-orbit coupling[22,28,29].

**Graphene-like band structure**
Below we focus on the Γ-valley region. Figure 2a shows the doping dependence of $R_{xx}$ at $E = 0$ and 0.1 V/nm. We observe resistance peaks around $\nu = 1/2, 2/3, 1, 1.5, 2$ and 2.2. The first three states are correlated insulators with diverging resistance in the zero-temperature limit. They are indicative of the presence of strong electron-electron interactions. The last three resistance peaks are below 10 kΩ and show a metallic temperature dependence (Fig. 4a). Because of strong interlayer coupling, the electric-field dependence of the Γ-valley states is generally weak[28] (except the correlated insulator at $\nu = 1$ which is beyond the scope of the current study).

We correlate the data in Fig. 2a with the doping dependence of Hall density, $n_H$, at $E = 0$ V/nm in Fig. 2b. The Hall density, $n_H = \frac{B_\perp}{eR_{xy}}$, is measured through the weak-field Hall resistance $R_{xy}$, where $e$ denotes the elementary charge, and $B_\perp$ (= 0.5 T), the out-of-plane magnetic field. As doping density increases, $n_H$ follows the doping density (lower dashed line) except near the correlated insulating states. The negative sign shows that the charge carriers are holes. Near $\nu = 1.5$ and 2.2, $n_H$ diverges and changes sign. Finally, $n_H$ vanishes at $\nu = 2$, and around it, again follows the doping density but with an offset of $2n_M$ (upper dashed line).

The observed doping dependence of $R_{xx}$ and $n_H$ supports the emergence of a graphene-like band structure (Fig. 1b,c). Specifically, the vanishing $n_H$ and the metallic temperature dependence of the resistance peak at $\nu = 2$ are consistent with the emergence of Dirac points at the $\pm \kappa$ points of the moiré Brillouin zone. The system is electron doped for $\nu < 2$ and hole doped for $\nu > 2$ in the vicinity of $\nu = 2$. Further, the sign change and amplitude reset of $n_H$, together with the resistance peak, near $\nu = 1.5$ and 2.2 are consistent with van Hove singularities in the DOS that originate from the saddle points in the band structure. The resistance is enhanced due to the large DOS[20]; the Fermi surface topology and the charge carrier type change abruptly, giving rise to a Hall density reset by $2n_M$. The deviation of the van Hove singularities, particularly the second, from the expected fillings ($\nu = 3/2$ and $5/2$) for an ideal graphene band structure[3] likely originates from mixing of the graphene-like band with the higher-energy moiré valence bands. The effect is expected to be stronger at higher filling factors.

**Massless Dirac fermions**
We demonstrate the emergence of massless Dirac fermions around $\nu = 2$ by their unique response to a magnetic field. Figure 3a shows $R_{xx}$ as a function of $\nu$ and $B_\perp$ at $E = 0.1$ V/nm. A Landau fan emerges from $\nu = 2$. In general, the Landau levels (LLs) are more visible for $\nu < 2$, where the effect of higher-energy moiré bands is less important. Figure 3b is the schematic Landau fan diagram extracted from Fig. 3a by tracing the resistance dips. At low fields, LLs with filling factor $\nu_{LL} = $ -10, -6, -2 and 2 are discernable. Additional LLs $\nu_{LL} = $ -4 and -8 emerge at higher fields, and $\nu_{LL} = \pm 1$, at even higher fields. At the same time, the $\nu = 2$ state becomes an insulator instead of a semimetal. The corresponding Hall resistance is shown in Extended Data Fig. 5, where quantized Hall resistance, $R_{xy} = \pm h/2e^2$, is observed for $\nu_{LL} = \pm 2$. Similar results are obtained from another device (Extended Data Fig. 6).

The anomalous LL sequence of $\nu_{LL} = $ -10, -6, -2 and 2 observed at low fields is a hallmark of the massless Dirac fermions[1,2]. The sequence of equidistant steps of four persists through the Dirac point, where charge carriers change from electrons to holes. The step size of four originates from the fourfold spin-valley degeneracy of the charge carriers[3]. The uninterrupted step of four from $\nu_{LL} = -2$ to $\nu_{LL} = 2$, together with the quantized Hall conductance at half integer of $4e^2/h$, reveals the existence of a LL at the

Dirac point or zero energy, which is shared by electrons and holes[1,2]. This is equivalent to a $\pi$ Berry's phase at the Dirac point[3]. At high fields, the emergence of additional LLs and an insulating state at $\nu = 2$ mimics breaking of the spin-valley degeneracy and quantum Hall ferromagnetism in graphene[43,44] and warrants further investigations.

Another hallmark of massless Dirac fermions is a cyclotron mass $m^*$ that depends on the square root of charge carrier density $n$ (Ref. [1,2]). We determine $m^*$ from the temperature dependence of Shubnikov-de Haas oscillations using the Lifshitz-Kosevich formula[45] (Methods). Figure 3c shows the doping dependence of $R_{xx}$ at $B_\perp = 4.2$ T and representative temperatures (only the electron side is shown). With increasing temperature, the oscillations decay more rapidly at higher electron densities, suggesting larger $m^*$. This is demonstrated in the inset for $\nu_{LL} = -2$ and $-6$, corresponding to $\nu = 1.93$ and 1.78. Figure 3d summarizes the density dependence of $m^*$, which can be described by $m^* = \frac{\hbar\sqrt{\pi n}}{v_F}$ (Ref. [1,2]) with fitting parameter $v_F \approx (2.5 \pm 0.5) \times 10^4$ m/s (shaded area). Its deviation from the $\sqrt{n}$ dependence at high densities is due to departure of the moiré band from the linearly dispersing structure. The extracted Fermi velocity is in reasonable agreement with the calculated $v_F \approx 4.1 \times 10^4$ m/s from the continuum model; the value is nearly two orders of magnitude smaller than the Fermi velocity ($\approx 1 \times 10^6$ m/s) in graphene.

**Relativistic Mott transition**
The significantly reduced Fermi velocity in AB-BA WSe$_2$ opens an opportunity to investigate the relativistic Mott transition driven by electron-electron interactions[6-15]. We examine the twist angle effect on transport near the Dirac point at $E = 0$ V/nm and under zero magnetic field. Figure 4a and 4b show the doping dependence of $R_{xx}$ at varying temperatures for a 3°-twisted and 2.5°-twisted sample, respectively. At 3°, we observe a metallic behavior, where the sample resistance increases with increasing temperature, for the entire doping range including the Dirac point and van Hove singularities. In contrast, at 2.5° we observe an insulating behavior near $\nu = 2$. Figure 4c summarizes the temperature dependence of $R_{xx}$ at $\nu = 2$ for different twist angles examined in this study. As twist angle decreases, the behavior evolves systematically from metallic to insulating. We estimate the charge gap of the insulating state by performing a thermal activation analysis of the data (Extended Data Fig. 7). Figure 4d shows that a finite gap develops at a critical angle of about 2.7° and the gap size increases with decreasing angle.

The observed insulating state at $\nu = 2$ in small twist angle samples cannot be explained by the single-particle band picture. The massless Dirac fermions are a consequence of the honeycomb lattice symmetry (Fig. 1a), which is preserved regardless of the twist angle and structural relaxations in the moiré length scale (Extended Data Fig. 2). Disorder effects near the Dirac point would only introduce electron and hole puddles[46] but cannot induce a transition to an insulating state due to Klein tunneling[20]. The observed semimetal-insulator transition is consistent with the spontaneous chiral symmetry breaking transition induced by the strong Coulomb repulsion between electrons[6-15]. We evaluate the interaction strength using the fine-structure constant or the Wigner-Seitz radius, $r_s = \frac{e^2}{2\epsilon\epsilon_0 h v_F}$ (Ref. [20]), in the upper axis of Fig. 4d, where $v_F$ is estimated from the continuum model band structure and $\epsilon \approx 4.5$ (Ref. [47]) and $\epsilon_0$ denote the dielectric constant of the medium (hexagonal boron nitride) and the vacuum permittivity, respectively. The transition occurs at $r_s \approx 13$. The value is comparable to but larger than the quantum Monte Carlo result of 4 for the half-filled Hubbard model on a honeycomb lattice[14]; part of the discrepancy may come from a lower estimate of $v_F$ by the continuum model.

**Discussion and conclusion**
We have demonstrated that the Γ-valley moiré valence band in twisted WSe$_2$ tetralayers realizes artificial graphene but with a fine-structure constant that can be tuned by varying twist angle. We have observed a spontaneous chiral symmetry breaking transition driven by electron-electron interactions when the twist angle is below about 2.7°. Several candidate ground states can break the chiral symmetry, including a Mott insulator[6,8,12,14] and charge-ordered states[10-12,15,18,19] that are stabilized by the onsite and extended Coulomb repulsions, respectively. Our initial study of the magnetoresistance and Hall response suggests that doping the $\nu = 2$ insulator results in an antiferromagnetic metal below a critical temperature and magnetic field (Extended Data Fig. 8). The result is compatible with the Mott scenario because the honeycomb Mott insulator is expected to be antiferromagnetic[6,8,12,14]. The Mott scenario is also consistent with the observed critical $r_s$ and the strongly screened extended Coulomb repulsions in our dual-gated device structure. Future studies with fine twist angle control will test whether the relativistic Mott transition is continuous with criticality belonging to the Gross-Neveu universality class, as theoretical studies predict[6,8,12,14].

**Methods**
**Device design and fabrication**
We fabricated twisted WSe$_2$ tetralayers from natural bilayers using the 'tear-and-stack' technique[48] and completed the devices with dual gates by the layer-by-layer dry transfer method[49]. Extended Data Fig. 1a-c show the device schematic (a) and optical images of two representative devices (b, c). Atomically thin flakes of WSe$_2$, hexagonal boron nitride (hBN) and graphite were exfoliated from bulk crystals (HQ Graphene) onto Si/SiO$_2$ substrates. Flakes of appropriate shape and thickness were picked up at about 70°C using a polymer stamp made of a thin layer of polycarbonate on a polypropylene-carbonate-coated polydimethylsiloxane block in the following sequence: hBN, top graphite gate (TG), hBN, bilayer WSe$_2$, bilayer WSe$_2$ with a twist, hBN, and bottom graphite gate (BG). The typical thickness of the hBN layers is about 10-15 nm. The finished stack was released by melting the polymer stamp at about 200°C onto Si/SiO$_2$ substrates with prepatterned platinum (Pt) electrodes. The contact gates (CG) and control gates (CoG) were then fabricated using electron-beam lithography, followed by palladium (Pd) deposition using an electron-beam evaporator.

The channel consists of the dual-gated region of the WSe$_2$ tetralayer by both the top and bottom gates. It is connected to the Pt electrodes through heavily doped WSe$_2$ regions by the contact gates directly above them. This approach allowed us to achieve relatively low metal-WSe$_2$ contact resistances (<100 kΩ) for low-temperature transport measurements at low doping densities and small electric fields[36,37]. The control gates were used to turn off the single-gated regions of WSe$_2$ so that they do not contribute to transport.

**Transport measurements**
The electrical transport measurements were performed in closed-cycle $^4$He cryostats (Oxford TeslatronPT and Cryomagnetic) down to about 1.5 K and in a dilution refrigerator (Bluefors LD250) down to about 20 mK in lattice temperature. A standard low-frequency (7–17 Hz) lock-in technique was employed to measure the sample resistances. The bias current and voltage drop were simultaneously recorded (Stanford Research SR830). The bias current was typically maintained at about 10 nA or below. The voltage was measured through a voltage pre-amplifier (Ithaco DL1201) with an input impedance of 100 MΩ.

**Twist angle and Landau level degeneracy calibration**
We investigated WSe$_2$ tetralayers with the relative twist angle between the two natural bilayers varying from 1.9° to 3.7°. The twist angle or the moiré density was calibrated using the resistance peaks at $\nu = 1$ and 2 and the period of the quantum oscillations observed under a perpendicular magnetic field. Details please refer to a related study on twisted bilayer WSe$_2$ (Ref. [37]). In short, we first acquired a resistance map under zero magnetic field as a function of the top gate and bottom gate voltages, and converted the gate voltages to $E$ and $\nu$ using the known hBN thickness (from atomic force microscopy measurements). We then obtained another resistance map at 12 T magnetic field as a function of $E$ and $\nu$ (Extended Data Fig. 4). We focus on the Landau levels under high electric fields, where the holes are pushed to the K/K' valley and the topmost/bottommost layer of the tetralayer; the Landau levels are non-degenerate in this layer- and valley-polarized region, similar to the case in hole-doped monolayer WSe$_2$ under large magnetic fields[50]. Using the known density difference $\frac{B_\perp}{\phi_0}$ between successive Landau levels in this region ($\phi_0 = \frac{h}{e}$ is the magnetic flux quantum), we can calibrate both $n_M$ and $\theta$. The non-degenerate Landau levels in this region also provides a calibration for the Landau level degeneracy near the Dirac point in Fig. 3.

**Extraction of the cyclotron mass**
The cyclotron mass was extracted from the temperature dependence of the Shubnikov-de Haas oscillation amplitude ($\Delta R_{xx}$) using the Lifshitz-Kosevich formula[45], $\Delta R_{xx} \propto \frac{\lambda(T)}{\sinh \lambda(T)}$ (inset of Fig. 3c). The amplitude $\Delta R_{xx}$ was obtained by normalizing $R_{xx}(T)$ at low temperatures to $R_{xx}(T)$ at high temperatures, where Shubnikov-de Haas oscillations are no longer discernable. The cyclotron mass $m^*$ is expressed through $\lambda(T) = \frac{2\pi^2 k_B T}{\hbar \omega_c}$, where $\omega_c = \frac{eB_\perp}{m^*}$ is the cyclotron frequency, and $k_B$ denotes the Boltzmann constant. Multiple Landau levels are discernable for magnetic field ranging from 2 T to 12 T. We focused on the oscillation amplitude of the $\nu_{LL} = \pm 2$ Landau levels since they are the most robust states (Fig. 3a,c) and persist over a wide temperature range. Under varying magnetic fields, the $\nu_{LL} = \pm 2$ Landau levels appear at different carrier densities, from which the density dependence of the cyclotron mass (Fig. 3d) is extracted through the temperature dependence.

**Band structure and DOS calculations**
The band structure of two natural WSe$_2$ bilayers with a small twist angle was calculated using the continuum model described in Ref. [28]. The moiré effect is assumed to appear only between the two inner layers. The Hamiltonian for the Γ-valley states is given by

$$H_{4L} = -\frac{\hbar^2 k^2}{2m_\Gamma} + \begin{pmatrix} \Delta_1(r) & \Delta_{12}(r) & 0 & 0 \\ \Delta_{12}^\dagger(r) & \Delta_2(r) & \Delta_{23}(r) & 0 \\ 0 & \Delta_{23}^\dagger(r) & \Delta_3(r) & \Delta_{34}(r) \\ 0 & 0 & \Delta_{34}^\dagger(r) & \Delta_4(r) \end{pmatrix}, \quad (1)$$

which is identical for both spin states. Here the first term represents the kinetic energy with $k$ and $m_\Gamma$ denoting, respectively, the crystal momentum and the effective mass of the Γ-valley holes. The diagonal elements in the second (potential energy) term are the intralayer potentials, $\Delta_1 = \Delta_4 = V_1$ for the two outer layers and $\Delta_l(r) = V_l^{(0)} + 2V_l^{(1)} \sum_{i=1,3,5} \cos(G_i \cdot r \pm \varphi)$ for the two inner layers ($l = 2, 3$). The off-diagonal elements describe interlayer tunneling, $\Delta_{12}(r) = \Delta_{34}(r) = V_{12}$ and $\Delta_{23}(r) = V_{23}^{(0)} + 2V_{23}^{(1)} \sum_{i=1,3,5} \cos(G_i \cdot r)$, where $G_i$'s are the moiré reciprocal lattice vectors. In the calculation we used the same parameters as in Ref. [28]: $(V_1, V_{12}, V_2^{(0)}, V_2^{(1)}, V_{23}^{(0)}, V_{23}^{(1)}) =$ (200, 184, -159, -8, 356, -9) meV, $\varphi = -0.17$ and $m_\Gamma = 1.0\, m_0$ (Extended Data Fig. 9), with $m_0$ being the electron rest mass.

To obtain the band structure, the Hamiltonian was diagonalized in the plane-wave basis at each momentum. The momentum cutoff was chosen to be at the 5$^{th}$ shell of the moiré Brillouin zone. To obtain the density of states, a histogram of energy eigenvalues was calculated at 106 discrete momentum points covering the first moiré Brillouin zone. The histogram was normalized by the number of points and the moiré unit cell area to convert to units of eV$^{-1}$nm$^{-2}$. The cumulative distribution function of the energy eigenvalues was used to determine the filling factor axis.

**Acknowledgements**
We thank Chaoming Jian for many helpful discussions.

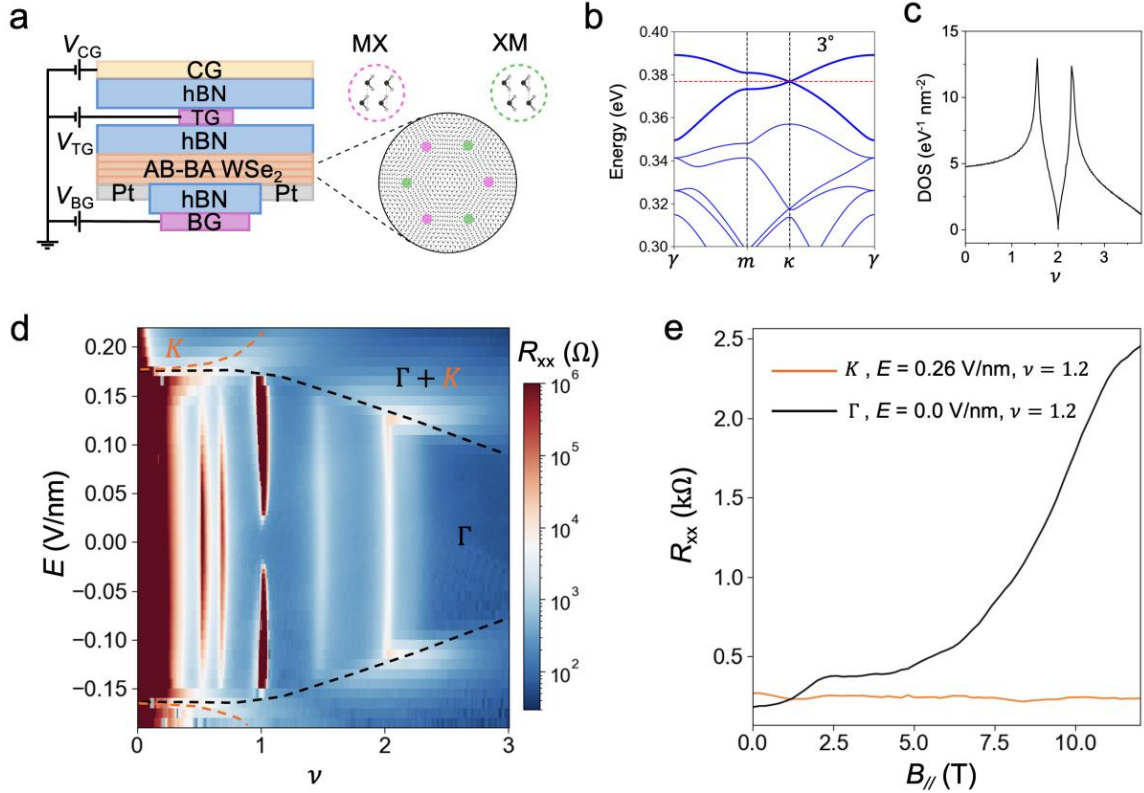

**Figure 1 | Γ-valley moiré bands in AB-BA WSe$_2$ tetralayers**. **a,** Schematic cross-section for the dual-gated AB-BA WSe$_2$ device contacted by Pt electrodes (grey). Both top (TG) and bottom (BG) gates are made of few-layer graphite electrodes (pink) and hBN dielectric (blue). The Pd contact gate (CG, yellow) turns on the Pt contacts. Two AB-stacked WSe$_2$ natural bilayers are twisted at a relative angle of 183 degrees (AB-BA stacking). At the interface, a honeycomb moiré lattice (inset) emerges with high symmetry points at the MX and XM sublattice sites (M=W; X=Se). **b,** Calculated band structure along the path $\gamma - m - \kappa - \gamma$ in the moiré Brillouin zone. The graphene-like bands are highlighted by thicker lines. The red dashed line marks the Fermi level at the Dirac point. **c,** Electronic density of states (DOS) obtained from the band structure as a function of the hole filling factor $\nu$. The two sharp peaks at $\nu \approx 1.5$ and $\nu \approx 2.3$ correspond to the van-Hove singularities in the band structure. **d,** Longitudinal resistance $R_{xx}$ as a function of $\nu$ and $E$ at zero magnetic fields and $T = 20$ mK (device 1). The orange and black dashed lines separate three distinct regions with different valley compositions for the moiré bands: Γ, $K$ and Γ + $K$ valleys in WSe$_2$. **e,** In-plane magnetoresistance in the Γ (black) and $K$ (orange) regions selected from **d**. The doping density and electric field are labeled. The $K$-(Γ-)valley holes show negligible (strong) in-plane magnetoresistance.

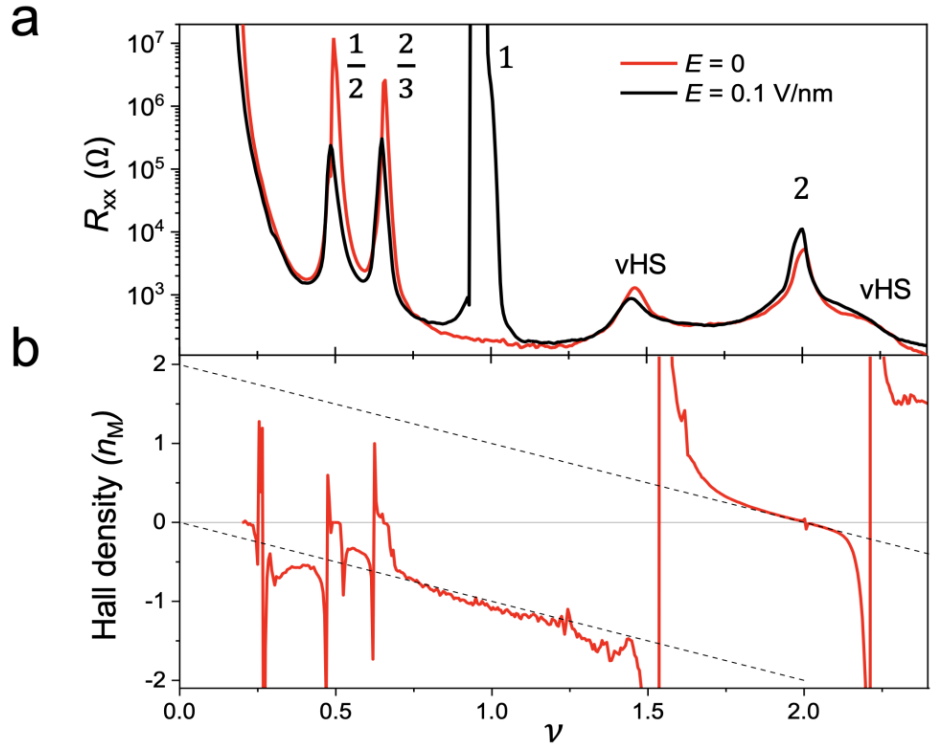

**Figure 2 | Graphene-like band structure revealed by transport measurements. a,** Line cuts of Fig. **1d** at two electric fields: $E = 0$ V/nm (red) and $E = 0.1$ V/nm (black). The resistance peaks at $\nu = 1/2$, $2/3$, $1$, $2$ and at the van Hove singularities (vHS) are labeled. **b,** The corresponding weak-field Hall density (measured at $B_\perp = 0.5$ T) as a function of $\nu$ at $E = 0$ V/nm. The Hall density is measured in units of the moiré density $n_M \approx 2.91 \times 10^{12}$ cm$^{-2}$. The dashed lines provide an eye guide to the doping densities with reference to $\nu = 0$ (lower) and $\nu = 2$ (upper). The Hall density crosses zero linearly at $\nu = 2$; it also changes sign at the vHS near $\nu = 1.5$ and $2.2$; the dependence is consistent with the DOS in Fig. **1c**.

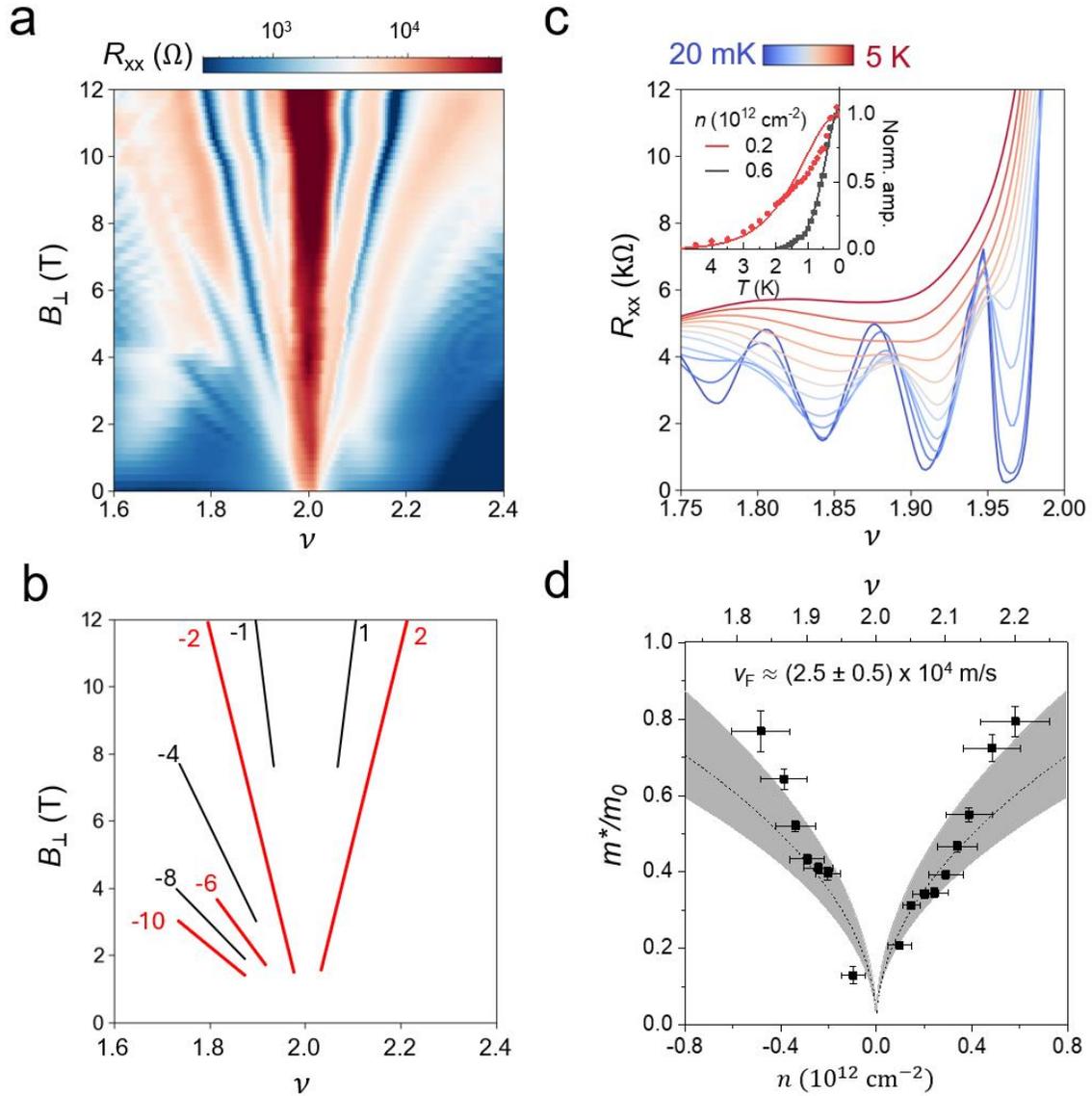

**Figure 3 | Massless Dirac fermions. a,** Longitudinal resistance $R_{xx}$ as a function of $\nu$ and $B_\perp$ at $E = 0.1$ V/nm and $T = 20$ mK (device 1). $R_{xx}$ is symmetrized under $B_\perp > 0$ and $B_\perp < 0$. **b,** A Landau fan centered at $\nu = 2$ constructed by tracing the resistance dips in **a.** At low magnetic fields, the Landau levels follow the sequence $\nu_{LL}$ =-10, -6, -2, 2 (red) for massless Dirac fermions with four-fold degeneracy. Only at higher fields that symmetry breaking ground states emerge at $\nu_{LL}$ =-8, -4, -1, 1 (black). **c,** Density dependent Shubnikov-de Haas oscillations at fixed $B_\perp = 4.2$ T and varying temperatures (from 20 mK to 5 K, blue to red). Only $B_\perp > 0$ data are shown. The inset shows the oscillation amplitude as a function of temperature at $n = 0.2 \times 10^{12}$ cm$^{-2}$ and $0.6 \times 10^{12}$ cm$^{-2}$. Solid lines are fits to the data using the Lifshitz-Kosevich formula. **d,** Quasiparticle effective mass extracted from **c** as a function of density. The effective mass follows the dependence $m^* = \frac{\hbar\sqrt{\pi n}}{v_F}$ (dotted line) with Fermi velocity $v_F \approx (2.5 \pm 0.5) \times 10^4$ m/s. The grey-shaded area represents the fitting uncertainty.

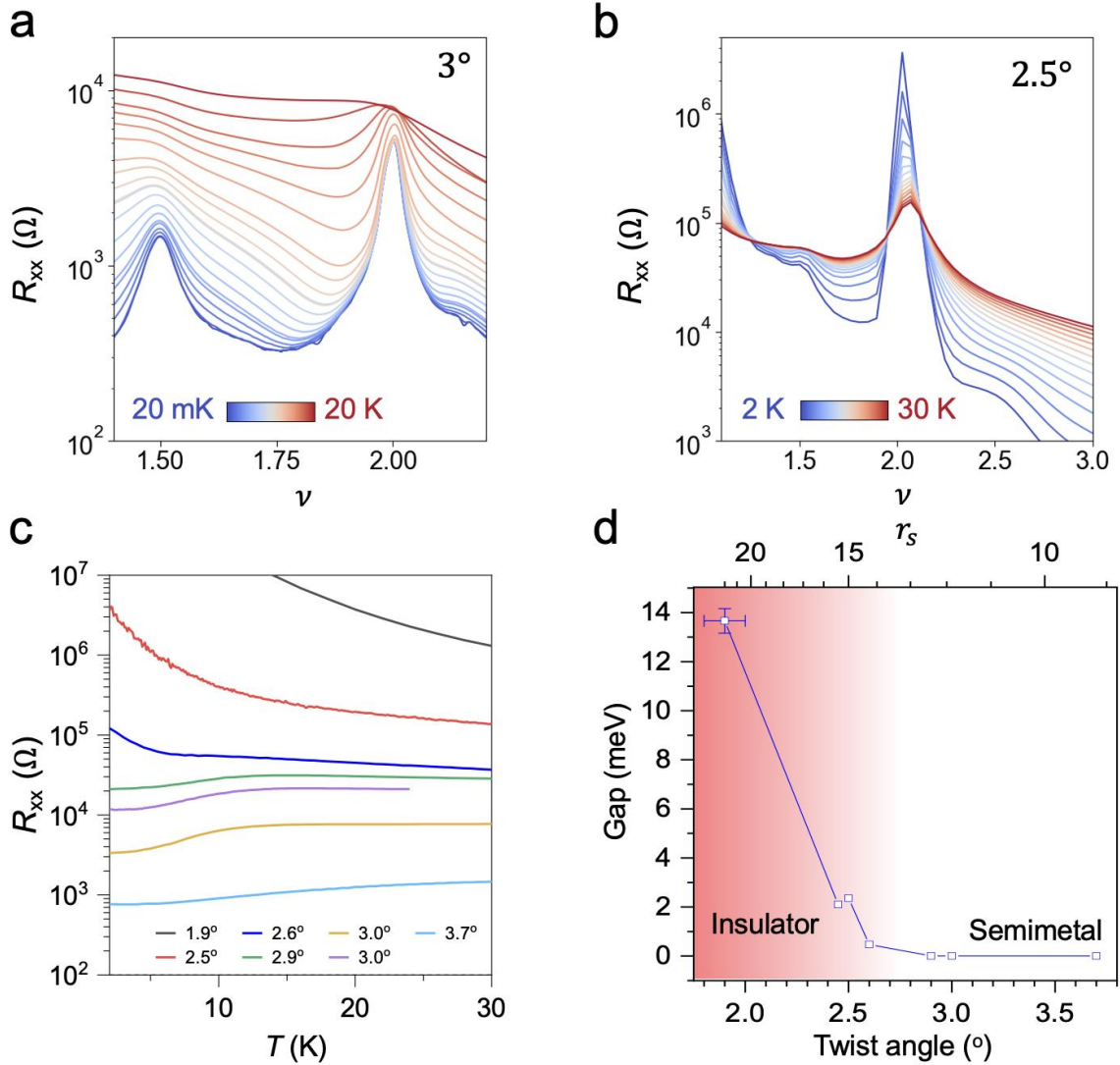

**Figure 4 | Relativistic Mott transition for Dirac fermions a,b,** Density dependence of $R_{xx}$ near $\nu = 2$ ($E = 0$ V/nm and $B_\perp = 0$ T) at varying temperatures for twist angle $\theta = 3^o$ (**a**) and $\theta = 2.5^o$ (**b**). **c,** Temperature dependence of $R_{xx}$ at $\nu = 2$ for devices with different twist angles from 1.9 to 3.7 degrees. **d,** Twist angle dependence of the charge gap extracted from **c**. The top axis shows the fine structure constant for Dirac fermions. A Mott insulator emerges at twist angles below about 2.7 degree (red-shaded region).